\documentclass[twocolumn,physrev,superscriptaddress]{revtex4-2}
\usepackage{amsmath,amssymb}
\usepackage{graphicx}
\usepackage{epstopdf}
\usepackage{color}
\usepackage{paralist} 
\usepackage{lineno} %for counting lines

\newcommand{\Higgs}{\mathcal{H}}
\newcommand{\MQ}{\mathsf{Q}}
\newcommand{\Ncal}{\mathcal{N}}
\newcommand{\urm}{\mathrm{U}}
\newcommand{\surm}{\mathrm{SU}}
\newcommand{\sprm}{\mathrm{Sp}}

\newcommand{\sorm}{\mathrm{SO}}

\newcommand{\Scal}{\mathcal{S}}
\newcommand{\Lcal}{\mathcal{L}}
\newcommand{\multiset}[1]{\{ \hspace*{-.35em} \{ #1  \} \hspace*{-.35em} \}}
\usepackage[normalem]{ulem}

\begin{document}
% \linenumbers %counting lines
\title{Decay and Fission of Magnetic Quivers}
\author{Antoine Bourget}
\affiliation{Universit\'e Paris-Saclay, CNRS, CEA, Institut de physique th\'eorique,  91191, Gif-sur-Yvette, France}
\author{Marcus Sperling}
\affiliation{Fakult\"at für Physik, Universit\"at Wien, 
Boltzmanngasse 5, 1090 Wien, Austria}
\author{Zhenghao Zhong}
\affiliation{Mathematical Institute, University of Oxford, 
Andrew Wiles Building, Woodstock Road, Oxford, OX2 6GG, UK}

\begin{abstract}
\noindent 
In exploring supersymmetric theories with 8 supercharges, the Higgs branches present an intriguing window into strong coupling dynamics. Magnetic quivers serve as crucial tools for understanding these branches. Here, we introduce the \emph{decay and fission} algorithm for unitary magnetic quivers. It efficiently derives complete phase diagrams (Hasse diagrams) through convex linear algebra. It allows magnetic quivers to undergo \emph{decay} or \emph{fission}, reflecting Higgs branch RG-flows in the theory. Importantly, the algorithm generates magnetic quivers for the RG fixed points and simplifies the understanding of transverse slice geometry with no need for a list of minimal transitions. In contrast, the algorithm hints to the existence of a new minimal transition, whose geometry and physics needs to be explored.
\end{abstract}

\maketitle

\section{Introduction}

The study of phases of quantum field theory (QFT), and transitions between these phases, is fundamental to our understanding of nature. One example is the Higgs mechanism, where a scalar field acquires a vacuum expectation value (VEV) that subsequently breaks the gauge symmetry \cite{englert1964broken,higgs1964broken,guralnik1964global,kibble1967symmetry}. Beyond its central role in the electroweak sector of the Standard Model, this mechanism can be observed in any gauge theory with charged scalar fields. It can further be generalized to intrinsically strongly coupled theories, which lack a Lagrangian description, but in this case a systematic understanding is still lacking, in part due to the extraordinary diversity of the landscape of QFTs. 

This can be remedied by adding simplifying assumptions. A particularly relevant playground in that respect is supersymmetry. Supersymmetric theories (SQFTs) encompass a very rich set of phenomena, from Standard Model-like theories to strongly coupled systems, while maintaining computational control. SQFTs with 8 supercharges in space-time dimension 3, 4, 5 and 6 generically possess a continuous space of vacua known as the Higgs branch, denoted $\Higgs$, parameterized by scalar fields (sitting only in the so-called hypermultiplets) when the field approach is available. The Higgs mechanism is then geometrized, and corresponds to hitting singularities in $\Higgs$. The phase diagram can then be identified with the structure of nested singularities \cite{Bourget:2019aer}. Mathematically, this so-called \emph{Hasse diagram} describes the finite stratification of the symplectic singularity $\Higgs$ into its partially ordered symplectic leaves. The latter can be understood as sets of (Higgs branch) VEVs that trigger distinct partial Higgs mechanisms.
The phase/Hasse diagram also encodes how SQFTs are related through deformations, tunings of gauge couplings, and RG-flows. 

The description of $\Higgs$ beyond the perturbative regime, which includes most conformal SQFTs, is challenging. Fortunately, a powerful technique has recently been introduced: one defines $\Higgs$ by means of an auxiliary combinatorial object called a \emph{magnetic quiver} (MQ)\cite{Cabrera:2018jxt,Cabrera:2019izd,Bourget:2019rtl,Cabrera:2019dob, Bourget:2020gzi,Bourget:2020asf,Bourget:2020xdz,Closset:2020scj,Akhond:2020vhc,vanBeest:2020kou,Giacomelli:2020gee,Bourget:2020mez,VanBeest:2020kxw,Closset:2020afy,Akhond:2021knl,Bourget:2021csg,vanBeest:2021xyt,Sperling:2021fcf,Nawata:2021nse,Akhond:2022jts,Giacomelli:2022drw,Hanany:2022itc,Fazzi:2022hal,Bourget:2022tmw,Fazzi:2022yca,Nawata:2023rdx, Bourget:2023cgs,DelZotto:2023nrb,LCLM:2023}. In the simplest cases, a MQ defines an auxiliary 3d $\mathcal{N}=4$ superconformal field theory (SCFT) whose Coulomb branch \footnote{The Coulomb branch of an SQFT with 8 supercharges is another maximal branch of the space of supersymmetric vacua. For 3d $\mathcal{N}=4$ the Coulomb branch is parameterized by VEVs of (dressed) monopole operators \cite{Borokhov:2002cg,Borokhov:2003yu} and is geometrically a symplectic singularity.} coincides by definition with $\Higgs$. It is known that a MQ encodes the Higgs phase diagram. 
An algorithm, dubbed \emph{quiver subtraction} \cite{Cabrera:2018ann,Bourget:2019aer,Bourget:2020mez,Bourget:2022ehw,Bourget:2022tmw}, has been developed over the last five years to extract this phase diagram under certain limiting conditions: 
\begin{compactitem} 
\item It holds for theories for which the class of phase transitions is already known. That is it requires as an external input the list of magnetic quivers for all elementary Higgsings (so-called \emph{isolated symplectic singularities}), which is still incomplete.
\item It provides very limited information on the
SQFTs at the end of the Higgs branch RG-flows.
\item It is computationally complex, it being an algorithm acting on the graphs underlying the quivers.
\end{compactitem}

This Letter introduces a powerful operation on quivers which addresses all the above problems: it provides the phase diagram of nested singularities of $\Higgs$ \emph{without relying on any input}, all the intermediate steps are well-defined, and the computation involves vectors, and not graphs. Moreover, the algorithm generates the magnetic quivers of SQFTs at the end of Higgs branch RG-flows; therefore, allowing to propose candidate SQFTs. In many circumstances, including further known data (e.g.\ string theory constructions, central charges, Coulomb branch spectra) then allows to identify a single candidate SQFT. In a companion paper \cite{Bourget:2024mgn}, we apply the algorithm to SQFTs in various dimensions and show perfect agreement with the literature, along with several new results. We also give there the details of the geometry of transverse slices, which are omitted here.

\section{Decay and Fission}

The algorithm can be described using analogies with nuclear reactions, applied to unitary quiver theories \footnote{In mathematics, a \emph{quiver}  is a directed graph, defined as a set of vertices connected by arrows. The \emph{double quiver} is obtained by adjoining the reverse arrow for each arrow --- thus, an undirected edge arises.

In physics, \emph{quivers} (in string theory \cite{Douglas:1996sw}) or \emph{mooses} (in particle physics \cite{GEORGI1986274}) are graphs representing the matter content of (usually supersymmetric) gauge theories. Quivers $\mathsf{Q}$ for 8 supercharges theories are undirected by convention. Expressing $\mathsf{Q}$ in 4 supercharges (e.g.\ a hyper becomes two chirals), $\mathsf{Q}$ becomes a double quiver.}. Let us unpack this statement: in this Letter, a quiver is an undirected graph, composed of nodes and edges, that encodes a 3d $\Ncal=4$ SCFT. It does so by specifying each node by the rank $k$ of a unitary $\urm(k)$ gauge group, and the (bi-fundamental) matter content by the edges in between nodes. A \emph{magnetic} quiver is a quiver for which emphasis is placed on its 3d $\Ncal=4$ Coulomb branch. Given a magnetic quiver $\MQ$, it can \emph{decay}, i.e.\ become lighter, or \emph{fission} into exactly two smaller quivers.  These processes are then repeated in all possible ways until one reaches \emph{stable} quivers, which can neither decay nor fission into non-trivial quivers. The decay and fission reaction diagram then coincides with the Higgs phase diagram of the theory of which $\MQ$ was a magnetic quiver. The rules specifying when decays and fissions occur are given next in Sec.~\ref{sec:math}. Before delving into the
details, we preview what these processes look like on concrete examples, and provide a rationale for our proposed algorithm.

\paragraph{Decay. }
During this transition, a given quiver (specified by shape and ranks) ``decays'' into a quiver of the same shape, but with reduced ranks. For example, 
\begin{align}
\raisebox{-.5\height}{
 \includegraphics[page=1]{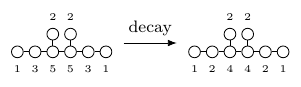}
 }
 \label{eq:example_decay}
\end{align}
which reflects the Higgs branch RG-flow of the 5d $\Ncal=1$ SCFT with low-energy effective description $\surm(5)_{1}$ gauge theory with $N_{\mathrm{AS}}=2$, $N_f=4$ to the SCFT fixed point of $\surm(4)_1$ with $N_{\mathrm{AS}}=2$, $N_f=4$.
During decay, the change in ranks (or equivalently, the dimension of the transition) can be any integer between one and the entire sum of the ranks. The latter is a terminal decay; i.e.\ the entire quiver decays to nothing. 

\paragraph{Fission. }
A given quiver can also ``fission'' into two quivers of the same shape, but with smaller ranks. For instance, the same quiver as above fissions as follows:
\begin{align}
\raisebox{-.5\height}{
 \includegraphics[page=1]{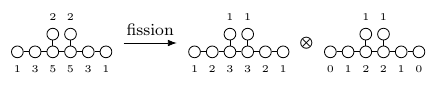}
 }
 \label{eq:example_fission}
\end{align}
which translates into the Higgs branch RG-flow of the $\surm(5)$ theory into a product of fixed points defined by $\surm(3)_1$ with $N_f=6$ and $\surm(2)$ with $N_f=4$.
In contrast to decay, the change in (sum of all) ranks during fission is always one; thus, these are 1-dimensional transitions. The $0$ nodes in \eqref{eq:example_fission} can be dropped, as done from now on.

\paragraph{Physical interpretation. }
The decay transition finds its origin in 3d mirror symmetry \cite{Hanany:1996ie,Gaiotto:2013bwa,Cabrera:2016vvv,Gu:2022dac}, in which the correspondence with Higgsing is manifest. The fission transition, on the other hand, can be understood as a generalization of adjoint Higgsing in theories with higher supersymmetry, like $\mathcal{N}=4$ SYM in 4d with gauge group $\mathrm{U}(k)$, which can be Higgsed to $\mathrm{U}(k') \times \mathrm{U}(k-k')$. In that case, the phases are labeled by partitions \cite{braden2014quantizations}, which fission obviously implements.

Our main result is that these two fundamental processes, when appropriately combined, cover all 8 supercharges theories in any dimension.

\paragraph{Proof of concept: Testing and Results. }  
Decay and fission consistently reproduces a large array of mathematical results such as nilpotent orbit closures, slices in the affine Grassmannian, or symmetric products. It can be derived from string theory in the (restricted) cases where Hanany-Witten brane setups and 3d mirror symmetry can be used. Physically, the algorithm reproduces the Higgs branch stratification of well-studied setups like partial closure of punctures in class $\Scal$ theories, or nilpotent Higgsings in 6d $\Ncal=(1,0)$ SCFTs. A host of such examples alongside with the new predictions offered by decay and fission are discussed in the companion paper \cite{Bourget:2024mgn}. In Sec.~\ref{sec:example}, the complete 5d $\Ncal=1$ example involving \eqref{eq:example_decay} and \eqref{eq:example_fission} is considered and direct validation from string theory arguments is given.

\section{The algorithm}
\label{sec:math}

In this section, we give the precise rules that govern the decay and fission processes introduced above. The \emph{underlying graph} of a quiver with $n$ unitary gauge nodes can be encoded into an adjacency matrix $A \in \mathrm{Mat}(n \times n ,\mathbb{Z})$. For $A_{i,j} = A_{j,i}$ with $i\neq j$, $A_{i,j} \geq 0$ is the number of (unoriented) links from node $i$ to node $j$; if $A_{i,j} >1$, the link is said to have multiplicity $A_{i,j}$. By convention $A_{i,i} = 2g-2$ where $g \geq 0$ is the number of loops on node $i$. If $A_{i,j} < A_{j,i}$ for $i\neq j$, we assume further that $\frac{A_{j,i}}{A_{i,j}} = \ell \in \mathbb{Z}_{>0}$; the edge between nodes $i$ and $j$ is then $\ell$-non-simply laced \footnote{This generalization was proposed in \cite{Cremonesi:2014xha}, in analogy with Dynkin diagrams. Each node can be given a \emph{length}, and the nodes with maximal length are called \emph{long}. Although this has no known Lagrangian description yet, it is motivated by brane systems with orientifolds \cite{Hanany:2001iy}, and the Coulomb branch has a working definition in terms of Hilbert series \cite{Cremonesi:2014xha} and is mathematically understood \cite{Nakajima:2019olw}.} and may have multiplicity $A_{i,j}=\frac{A_{j,i}}{\ell} \geq 1$. The \emph{quiver} itself is specified by its rank vector $K \in \mathbb{N}^n$. Given two rank vectors $K_{1,2} \in \mathbb{N}^n$, we say $K_1 \leq K_2$ if $K_2 -K_1 \in \mathbb{N}^n$. 

Finally, we need to introduce the essential notion of \emph{good} quivers. Our approach adopts the following concise, algorithm-friendly definition \footnote{Condition (i) means that every single unitary gauge node is good, i.e.\ having enough charged matter fields. Condition (ii) eliminates quivers that are a single $\urm(1)$ node (plus some adjoint hypermultiplets). Condition (iii) detects moduli spaces of instantons on $\mathbb{C}^2$, which come with a free hypermultiplet and therefore have two monopole operators that saturate the unitarity bound \cite{Cremonesi:2014xha}.\\
With our definition, an affine Dynkin quiver for the simple algebra $\mathfrak{g}$ wherein all ranks equal the $k$-fold of the dual Coxeter labels, which we denote $k \cdot \mathsf{Q}(\hat{\mathfrak{g}})$, is good.
} based solely on $(A,K)$: \begin{compactenum}[(i)]
     \item $AK \geq 0$,
     \item the sum of the entries of $K$ is at least 2, and
     \item there is no long node index $i$ with $K_i = 1$ such that $A \tilde{K} = 0$, with $\tilde{K}_j = K_i - \delta_{i,j}$.
 \end{compactenum}
Intuitively, a quiver is called good \cite{Gaiotto:2008ak} if all monopole operators \cite{Borokhov:2002cg,Borokhov:2003yu} in the 3d $\mathcal{N}=4$ theory have R-charge $\geq 1$ (above the unitarity bound). 

\paragraph{Partial Order. }
Given a good quiver $(A,K)$, the finite set of decay and fission products is defined by
\begin{equation}
 \mathcal{V} = \{  K' \in \mathbb{N}^n \; | \;  K' \leq K \, , (A, K') \textrm{ good}  \}  \,.
    \label{eq:decay_fission_products}
\end{equation}
To implement fission, we then assemble the products in all possibles ways: with $ \mathcal{V}_0 = \{ K' \in \mathcal{V} \; | \; K' \neq 0 \}$, define $L_0 = \{ \emptyset \}$ and 
\begin{align}
\label{eq:Lm}
    L_m = \{ \multiset{ K'_1 , \dots , K'_m  }  \;| \;   &\forall\; 1 \leq j \leq m \, , \, K'_j \in \mathcal{V}_0   \\ &\textrm{ and } K'_1 + \dots + K'_m \leq K \}  \notag 
\end{align}
for $m \geq 1$. Elements of $L_m$ are the multisets (i.e.\ with repetitions allowed) of $m$ vectors of $\mathcal{V}_0$ whose sum is $\leq K$. Finally, we write $L = \bigcup_{m \in \mathbb{N}} L_m$. 
This is the (finite) set of vertices of our Hasse diagram, which correspond to the symplectic leaves of the 3d $\Ncal=4$ Coulomb branch of the initial quiver $(A,K)$. For $\Lcal \in L$, the unique $m \in \mathbb{N}$ such that $\Lcal \in L_m$ is called the \emph{length} $\mathrm{length}(\Lcal)$ of $\Lcal$.
We also denote by $\Sigma \Lcal \in \mathbb{N}^n$ the sum of the elements of $\Lcal$. 

Next, we define a partial order on $L$. Let $\Lcal_1 , \Lcal_2 \in L$. We write $\Lcal_2 \rightsquigarrow \Lcal_1$ if   
\begin{compactitem}
    \item $| \mathrm{length}(\Lcal_1) - \mathrm{length}(\Lcal_2) | \leq 1$,  
    \item $\mathrm{length}(\Lcal_1 \cap \Lcal_2) \geq  \mathrm{length}(\Lcal_1) - 1 $, and 
    \item $\Sigma \Lcal_1 \geq \Sigma \Lcal_2$.
\end{compactitem} \noindent
The relation $\rightsquigarrow$ is reflexive and antisymmetric, but not transitive in general. Let us denote by $\succcurlyeq$ its transitive closure, i.e.\ $\Lcal_1 \succcurlyeq \Lcal_2$ if there exist a chain $\Lcal_1 \rightsquigarrow \dots \rightsquigarrow \Lcal_2$. This is a partial order relation. We claim that $(L, \succcurlyeq)$ coincides with the poset of symplectic leaves in the 3d $\Ncal=4$ Coulomb branch of the quiver $(A,K)$.

\paragraph{Elementary transitions. } 
The previous paragraph has shown how to associate to any quiver $(A,K)$ a poset of (multisets of) quivers. The partial order can be depicted using a Hasse diagram, in which elements of the poset are represented as points, and lines are drawn between \emph{adjacent} elements: there is a line between $\Lcal_1 \succcurlyeq \Lcal_2$ if there is no leaf $\Lcal_3 \neq \Lcal_1 , \Lcal_2$ satisfying $\Lcal_1 \succcurlyeq \Lcal_3 \succcurlyeq \Lcal_2$. This has a physical and a geometric incarnation. Physically, this plays the role of a Higgs branch phase diagram, in which lines represent \emph{elementary phase transitions}. Geometrically, we are in presence of a conical symplectic singularity, and lines correspond to \emph{minimal degenerations} \cite{Kraft1982,fu2017generic}. The degeneration/transition type is then indicated next to the line between two adjacent points in the Hasse diagram; see Fig.~\ref{fig:algorithm} for example.

It is of both physical and mathematical relevance to classify these elementary transitions. Interestingly, our algorithm, originally designed to compute the phase diagram for a particular theory, can be twisted to achieve this goal. Namely, search systematically through the set of all magnetic quivers, and identify those which have a phase diagram that consists of exactly two phases; the quiver then serves as a signature for that phase transition. A proof of principle is provided in Sec.~\ref{sec:perspectives}. 

\section{A complete example}
\label{sec:example}
\begin{figure}[t]
    \centering
    \scalebox{.9}{
    \includegraphics[page=1]{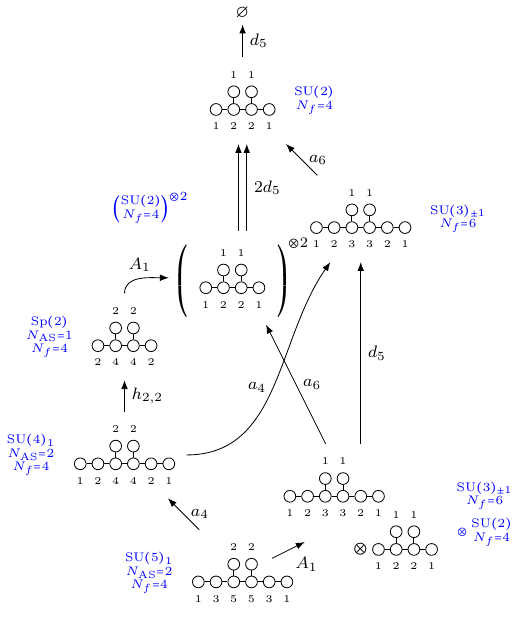}
    }
    \caption{Decay and fission algorithm for the 5d $\Ncal=1$ SCFT with $\surm(5)_1$ gauge theory description plus $N_{\mathrm{AS}}=2$ anti-symmetric and $N_f=4$ fundamental matter fields. The algorithm generates the entire Hasse diagram by constructing for each leaf one or more MQs. The minimal degenerations are labeled next to the arrows and can be obtained in a subsequent step, detailed in \cite{Bourget:2024mgn}. The corresponding 5d $\Ncal=1$ theories are indicated in blue.}
    \label{fig:algorithm}
\end{figure}

To illustrate the decay and fission algorithm, we consider the magnetic quiver at the bottom of Fig.~\ref{fig:algorithm} for the SCFT-fixed point of 5d $\Ncal=1$ $\surm(5)_{1}$ gauge theory with $N_{\mathrm{AS}}=2$ antisymmetric and $N_f=4$ fundamental hypermultiplets, and Chern-Simons level $1$ \cite{VanBeest:2020kxw}. Despite the low-energy effective description, the emergence of massless (gauge) instantons at the SCFT-fixed point means that the phase diagrams at the fixed point and away from it differ drastically. This is because the Higgs branch moduli at the fixed point include these instanton operators as well. The lack of a Lagrangian description at the fixed point implies that semi-classical approaches are insufficient. In contrast, the decay and fission algorithm outputs the phase diagram for the full ``quantum'' Higgs branch, as shown in Fig.~\ref{fig:algorithm}. The starting point can undergo two distinct processes: the decay~\eqref{eq:example_decay} and the fission~\eqref{eq:example_fission}. The resulting MQs readily allow the identification of the SCFTs which reside at the end of the Higgs branch RG-flows, due to the comprehensive catalogs of MQs for 5d theories \cite{Cabrera:2018jxt,Bourget:2020gzi,vanBeest:2020kou,VanBeest:2020kxw,Akhond:2020vhc}. Therefore, the decay and fission algorithm allows to construct the Higgs branch RG-flow diagram, as indicated in blue in Fig.~\ref{fig:algorithm}.

Physically, the 5d $\Ncal=1$ theory can be realized in Type IIB superstring theory by a 5-brane web \cite{Aharony:1997ju,Aharony:1997bh} with two O$7^-$ orientifolds. 
Crucially, the 5-brane web of the $\surm(5)$ theory with $N_{\mathrm{AS}}=2$ requires two fractional NS5 branes on the orientifolds \cite{Bergman:2015dpa};  the $N_f=4$ fundamental hypermultiplets are constructed by adding D7 branes. Here, we use the classical brane system to gain insights; the quantum-mechanical brane web is obtained after resolving O$7^-$ planes. The latter gives rise to the MQs at the conformal fixed point.

\begin{figure}[t]
    \centering
    \scalebox{.9}{
    \includegraphics[page=1]{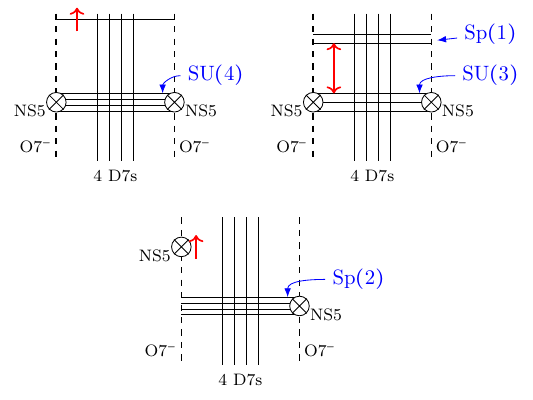}
    }
    \caption{Higgs branch motions in the (classical) 5-brane web with O$7^-$ orientifolds; wherein red arrows indicate which separation induces VEVs. Top Left: Decay $\surm(5)\to \surm(4)$ via moving a gauge 5-brane off to infinity along the flavor 7-branes. Top Right: Fission $\surm(5)\to \surm(3) \times \sprm(1)$ via splitting the stack of gauge branes into a stack of 2 gauge branes (away from the fractional NS5s) and 3 gauge branes (remaining between the NS5s). Bottom: Decay $\surm(4) \to \sprm(2)$ via moving a fractional NS5 off to infinity along the orientifold.}
    \label{fig:branes}
\end{figure}

The 5-brane web allows for two distinct Higgs branch transitions: firstly, a ``decay'' of the $\surm(5)$ theory to the $\surm(4)$ theory. This is realized by moving a single D$5$ brane (a gauge 5-brane) off to infinity (along the flavor 7-branes), see Fig.~\ref{fig:branes} (Top Left). Secondly, the fission is enabled by the presence of two O$7^-$ orientifolds (and the fractional NS5s). To see this, suppose that the gauge 5-branes all terminate on both fractional NS5 branes. One possible brane motion along the Higgs branch is to move one of the factional NS$5$ along the O$7^-$. However, this is only allowed if the total number of gauge branes is even, as required by the orientifold projection. Thus, the fractional NS5s cannot be moved for an odd number of gauge 5-branes. Instead, for any number $N$ of gauge 5-branes, one can ``fission'' the stack into $2\ell$ and $(N{-}2\ell)$, for $\ell \in \{1, 2,\ldots, \lfloor \frac{N}{2} \rfloor \}$. The stack of $2\ell$ branes can be moved away from the fractional NS5s, along the O$7^-$; while the stack of $(N{-}2\ell)$ branes remains suspended between the fractional NS5s, see Fig.~\ref{fig:branes} (Top Right). The latter subsystem then realizes an $\surm(N-2\ell)$ world-volume gauge theory with $N_{\mathrm{AS}}=2$ and some fundamental hypermultiplets. The former 5-brane subweb is $2\ell$ 5-branes intersecting two O$7^-$ planes (without any fractional NS5s) and some $N_f$ flavor 7-branes. This is the T-dual of $2\ell$ D4 branes inside a stack of $N_f$ coincident D8s in the presence of a single O$8^-$ --- the brane system realizing the moduli space of $\ell$ $\sorm(2N_f)$-instantons \cite{Witten:1995gx,Douglas:1995bn}. Therefore, the stack of $2\ell$ 5-branes between the two O$7^-$ planes together with the flavor 7-branes yields an $\sprm(\ell)$ world-volume theory with $N_{\mathrm{AS}}=1$ and some fundamental matter.  In the case of Fig.~\ref{fig:algorithm} with $N=5$, $\ell=1$, the Higgs branch RG-flow takes the $\surm(5)$ theory to a product of an $\surm(3)$ and an $\surm(2)\cong \sprm(1)$ theory.

The decay from the $\surm(4)$ to the $\sprm(2)$ theory is realized by moving one of the fractional NS5 branes along one of the O$7^-$ branes off to infinity. This follows, as the difference between an $\surm(4)$ brane web with $N_{\mathrm{AS}}=2$ and an $\sprm(2)$ brane web with $N_{\mathrm{AS}}=1$ is whether there are two or one fractional NS5 branes \cite{Bergman:2015dpa}, respectively, see Fig.~\ref{fig:branes} (Bottom).

We now turn to the physics of instantons to cross check parts of the Hasse diagram in Fig.~\ref{fig:algorithm}. The Higgs branch of the 
$\sprm(2)$ with $N_{\mathrm{AS}}=1$ and $N_f=4$ SCFT is the moduli space of two $\sorm(10)$-instanton on $\mathbb{C}^2$, with magnetic quiver $2 \cdot \mathsf{Q}(\hat{D}_5)$ (recall footnote [46]). To see this, view instantons as realized as $k$ D$p$ branes within a stack of coincident D$(p{+}4)$ branes \cite{Witten:1995gx,Douglas:1995bn}, in the presence of a suitable O$(p{+}4)$ orientifold. The MQ here is $k \cdot \mathsf{Q}(\hat{\mathfrak{g}})$ and the orientifold type determines $\mathfrak{g}$. The only possible Higgs branch motions are to split the branes into two stacks of $(k{-}\ell)$ and $\ell$ coincident branes. The magnetic quiver consequently fissions as $k \cdot \mathsf{Q}(\hat{\mathfrak{g}}) \rightarrow (k{-}\ell) \cdot \mathsf{Q}(\hat{\mathfrak{g}}) \otimes \ell \cdot \mathsf{Q}(\hat{\mathfrak{g}})$. 
Only if one of the stacks is a single D$p$ brane another Higgs branch motion arises: the D$p$ dissolves in the D$(p{+}4)$ stack and realizes the instanton moduli. From the MQ view point, $1 \cdot \mathsf{Q}(\hat{\mathfrak{g}})$ decays completely.    

The remaining transitions in Fig.~\ref{fig:algorithm} can be analyzed with similar arguments.

\begin{figure}
    \centering
    \scalebox{0.9}{
    \includegraphics[page=1]{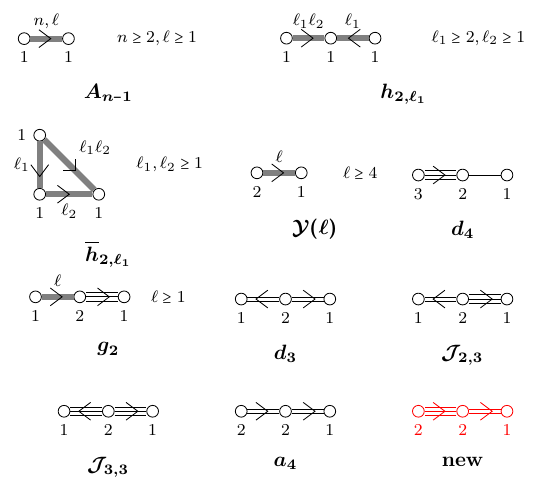}
    }
    \caption{List of all 2- and 3-nodes good unitary quivers that give rise to elementary Higgsing transitions. We use notations from \cite{Kraft1982,Bourget:2021siw,Bourget:2022tmw}. It includes the recently introduced slices $\mathcal{Y}(\ell)$ \cite{bellamy2023new}, and a new transition, in red.}
    \label{fig:3nodes}
\end{figure}

\section{Perspectives}
\label{sec:perspectives} 

The decay and fission algorithm places particular emphasis on quivers that cannot decay or fission into other non-trivial quivers. 
These stable quivers lead to 3d $\mathcal{N}=4$ Coulomb branches 
\cite{Nakajima:2015txa,Nakajima:2019olw} characterized by isolated conical symplectic singularities (ICSS). 
These can be seen as the geometric incarnation of elementary phase transitions on Higgs branches, in addition to offering interest to mathematics in their own right. This simple observation offers a well-defined perspective to list \emph{all} ICSS that can be realized as Coulomb branches of unitary quivers, reducing the problem to convex linear algebra, as stated in Sec.~\ref{sec:math}. As a proof of concept, we implemented this principle on all quivers with up to three nodes. The resulting list is presented in  Fig.~\ref{fig:3nodes}. Physically, this covers a surprisingly vast array of theories: these quivers appear within Higgs branches of theories ranging from 4d \cite{Bourget:2020asf} to 6d SCFTs \cite{Bourget:2022tmw}.
Remarkably, our exploration adds one new element (highlighted in red), corresponding to a new ICSS of complex dimension 8 with isometry $\mathfrak{so}_5$. We predict that the associated phase transition should show up in some yet to be discovered SCFTs. 

The above exploration will be carried out more generally to arbitrary size quivers in a future work, thereby charting a significant corner of the landscape of possible Higgs transitions. Combined with an extension to decorated quivers, we expect to be able to describe every single minimal degeneration using the Coulomb branch construction. 
However, the set of unitary quivers does not cover all possible (physically motivated) 3d $\Ncal=4$ quiver gauge theories, albeit it is a vast class of pivotal importance.  

Quivers involving gauge groups $\surm(n)$, $\sorm(n)$, $\sprm(n)$, or different matter field representations, (e.g.\ higher charge or non-fundamental representations), are not covered in this Letter, even though they appear prominently as magnetic quivers for higher-dimensional theories (see \cite{Cabrera:2019dob,Bourget:2020gzi} and subsequent works). To apply decay and fission to them, one needs to extend the combinatorial description of equations \eqref{eq:decay_fission_products} and \eqref{eq:Lm} and characterize the partial order, guided by known field and string theory constructions. 

Finally, we can further adapt our algorithm towards quiver growth. This adaptation, combined with our deep understanding of elementary transitions, paves the way for a bootstrap method; it opens possibilities for classifying theories based on their Higgs branch geometry, thereby complementing the Coulomb branch approach of \cite{Argyres:2015ffa,Argyres:2016xua}. 

\medskip\noindent {\bf Acknowledgements}
We thank Christopher Beem, Simone Giacomelli,   Julius Grimminger, Amihay Hanany,  Patrick Jefferson, Monica Kang, Hee-Cheol Kim, Sung-Soo Kim, Craig Lawrie, Lorenzo Mansi, Carlos Nunez, Matteo Sacchi, Sakura Sch\"afer-Nameki for many discussions on this topic. AB is partly supported by the ERC Consolidator Grant 772408-Stringlandscape.
MS is supported by the Austrian Science Fund (FWF), START project ``Phases of quantum field theories: symmetries and vacua'' STA 73-N. MS also acknowledges support from the Faculty of Physics, University of Vienna.
ZZ is supported by the ERC Consolidator Grant \# 864828 ``Algebraic Foundations of Supersymmetric Quantum Field Theory'' (SCFTAlg). 
Part of this work was realized at the Workshop on Symplectic Singularities and Supersymmetric QFT (July 2023, Amiens).

\bibliographystyle{apsrev4-2}
\bibliography{bibli}
\end{document}